\documentclass[twocolumns,fleqn,usenatbib]{mnras}

\usepackage[utf8]{inputenc}
\usepackage{scalefnt}
\usepackage{mathtools}
\usepackage{txfonts}
\usepackage{natbib}

\usepackage{graphicx}	
\usepackage{amsmath}	
\usepackage{amssymb}	

\title[ISRF and astrophysical ices]{The role of external FUV irradiation in the survival of astrophysical ices in Elias 29}

\author[Rocha \& Pilling]{
W. R. M. Rocha,$^{1,2}$\thanks{E-mail: willrobson88@hotmail.com}
S. Pilling,$^{1,3}$
\\
$^{1}$Instituto de Pesquisa e Desenvolvimento (IP\&D), Universidade do Vale do Paraíba, Av. Shishima Hifumi 2911, CEP 12244-000, São José dos Campos, SP, Brazil\\
$^{2}$SUPA, School of Physics \& Astronomy, University of St. Andrews, North Haugh, St. Andrews KY16 9SS, UK\\
$^{3}$Departamento de Física, Instituto Tecnológico de Aeronáutica, ITA - DCTA, Vila das Acácias, São José dos Campos, CEP 12228-900 SP, Brazil
}

\date{Accepted XXX. Received YYY; in original form ZZZ}

\pubyear{2015}

\begin{document}
\label{firstpage}
\pagerange{\pageref{firstpage}--\pageref{lastpage}}
\maketitle

\begin{abstract}
The survival of astrophysical ices in the star-forming regions depends on the suitability of temperature, density and radiation conditions. In this paper, the role of Interstellar Radiation Field (ISRF) on ices in Elias 29 is addressed. This object is the most luminous protostar in $\rho$ Oph E molecular cloud, and is surrounded by many Young Stellar Objects distant only a few arcminutes. In addition, other two brightness BV stars (S1 and HD 147889) enhance the external irradiation in Elias 29. This study was carried out by using the Monte Carlo radiative transfer code RADMC-3D assuming an internal and external irradiation. As result, we found that HD 147889 dominates the ISRF, instead of the closest protostars, and contributes to enhance the external irradiation in 44 times the standard value. Furthermore, remarkable effects are observed in FIR spectrum, as well as in near-IR image. Additionally, the snowline positions of volatile compounds, such as CO, O$_2$, N$_2$ and CH$_4$ are redefined to a toroidal-shaped morphology in the envelope, with low FUV flux (10$^{-7}$ ergs cm$^{-2}$ s$^{-1}$). In such scenario, the formation of complex molecules as result of hydrogenation or oxygenation of volatile species are expected to be severely affected. 

\end{abstract}



\section{Introduction}
The gravitational collapse of molecular clouds is followed by hierarchical fragmentation that breaks up a large molecular cloud into smaller cores to form low-mass stars \citep{Larson1978, Klessen1998}. In such a scenario of star-forming region, it is expected that all Young Stellar Objects (YSOs) are externally heated by neighborhood protostars, once they increase the FUV Interstellar Radiation Field (ISRF) by a few orders of magnitude. The standard ISRF might be defined as function of UV Draine field \citep{DraineBertoldi1996, Rollig2007} - $\chi_{ISRF}$ = 1) integrated between 91.2 – 205 nm (FUV regime) to ensure coverage important photochemical processes such as photodissociation \citep{vanDishoeck2006}. The numerical integration for $\chi_{ISRF}$ = 1 yields a  flux of 1.921 $\times$ 10$^8$ cm$^{-2}$ s$^{-1}$.
 
\citet{Jorgensen2006} discuss some effects of strong external irradiation on protostellar envelopes in Orion cloud. The authors report that ISRF was increased to 10$^3$ - 10$^4$ times the standard field, and consequently the outer envelope of protostars were heated above the desorption temperature of CO, i.e. larger than 25 K. Other regions, such as Corona Australis (CrA) were considered in \citet{LindbergJorgensen2012}. Using mainly H$_2$CO rotational diagrams, they found that outer envelopes in Cr A are heated up to 60 K by a closer Herbig Be star. This same region was probed later on by \citet{Lindberg2014}, employing ALMA observations. As conclusion, they state that evaporation of CO ice due to high temperature prevents the formation of Complex Organic Molecules (COMs) in CrA cloud.
    
Photodissociation regions (PDRs) can also be predicted by measuring the strength of ISRF using fine-structure emission lines such as OI (63 and 145 $\mu$m) and CII (158 $\mu$m) at FIR wavelenghts. \citet{Je2015} observed the Class I protostar GSS30-IRS1 located in Ophiuchus (Oph) cloud by using the Photodetector Array Camera and Spectrometer (PACS) on-board the Herschel Space Observatory. They report an extended CII emission, that probes a PDR region. The total CII intensity revels that standard ISRF is enhanced up to 20 in units of Habing Field (G$_0$). Indeed, the CII strong extended emission in $\rho$ Oph was previously described in \citet{Liseau1999} as result of an external irradiation of around 10 - 120 G$_0$ originated in B2V star HD 147889. The origin of these emissions might not be related with protostars itself, but rather can probe PDRs in outer parts of molecular clouds that hosts very embedded objects. The outer envelope, however, would be addressed by using H$_2$CO and c-C$_3$H$_2$ emission lines as suggested by \citet{Lindberg2017}. The authors also estimated that temperature of outer envelope in Oph cloud (excluding the upper limits) varies between 20 - 50 K, due to influence of two B-type stars: S1 and HD 147889.

Among the YSOs formed in Ophichus cloud, Elias 29 is a Class I object, placed at $\rho$ Oph E ($\alpha$(J2000) = 16h27'09.42'' and $\delta$(J2000) = -24$^{\circ}$37'21.1''), whose distance and luminosity are respectively 120 pc and 16.5 L$_{\odot}$ \citep{RochaPilling2015}, hereafter named Paper I. As suggested in Paper I, the frozen molecules in Elias 29 have experienced a chemical evolution driven by Cosmic Rays (CR), given the prominent absorption bands between 5 - 8 $\mu$m. However, many other objects formed in the vicinity of Elias 29, might play an important role for the survival of astrophysical ices in its envelope. In this paper, the photochemical processes such as photodesorption and photodissociation in Elias 29 are discussed, and consequently the half-life time of Complex Organic Molecules formed by interstellar irradiation. 

This paper is structured as follows: Section 2 constrains the external irradiation in Elias 29 and introduce the physical parameters for the continuum radiative transfer simulation using the RADMC-3D code. In Section 3, are shown the results and discussions relative to the 0.1 – 10000 $\mu$m SED, near-IR image, as well as the temperature profile at the outer envelope. Section 4, finally summarizes our conclusions.

The main conclusions are summarized in next.

\section{Methodology}
It is presented in this section how the external UV radiation field in Elias 29 was constrained from the observational data and simulated by a radiative transfer model with the RADMC-3D\footnote{http://www.ita.uni-heidelberg.de/~dullemond/software/radmc-3d/} code.  

\subsection{External irradiation in the protostar Elias 29}
A detailed study concerning the external irradiation in $\rho$ Ophiuchi cloud was firstly shown by \citet{Liseau1999}, although any specific statement about Elias 29 is not presented. According to authors, this cloud is surrounded by two potential sources of UV: HD 147889 (B2 V) and S1 (B3 V), although they conclude that HD 147889 dominates the radiation field in this scenario. S1 emission, on the other hand, only dominates in its neighborhood, i.e. $\rho$ Oph A as described in \citet{Liseau2015} and \citet{LarssonLiseau2017}. 

\begin{figure*}
\centering
\includegraphics[height=10cm]{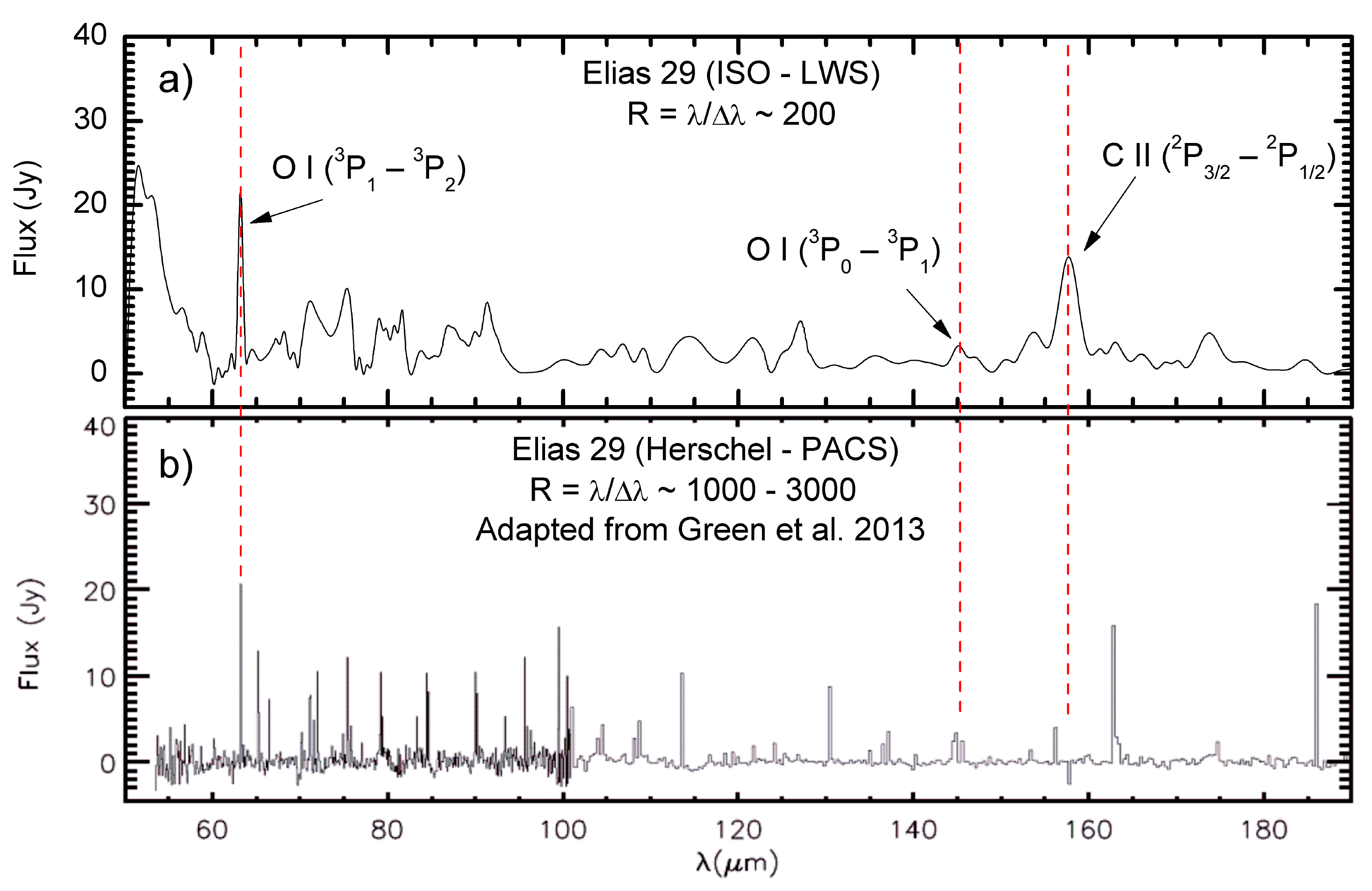}
\caption{Continuum-subtracted spectrum of Elias 29 in FIR. The fine-structure emission lines are indicated by arrows. Panel (a) shows an observation from ISO, whereas Panel (b) the spectrum obtained with Herschel Space Observatory adapted from \citet{Green2013}}
\label{emission}
\end{figure*}

The high illumination coming from HD 147889 produces a PDR-type emission at edges of $\rho$ Oph cloud \citep{Liseau1999, Ceccarelli2002, Green2016} given the presence of CII [158 $\mu$m] line detected with ISO-LWS between 45 - 195 $\mu$m \citep{Clegg1996} as shown in Figure \ref{emission}a. This line, however, was detected in absorption by the PACS instrument on-board the Herschel Space Observatory, with resolution power R $\sim$ 1000 - 3000 as reported by \citet{Green2013} and shown in Figure \ref{emission}b. The reason is because Herschel effectively chops back and forth every 1/8 of a second between the target (Elias 29) and an off position a few arcminutes away ($\sim$ 3') and since the [CII] was in both positions, it is subtracted off (private communication with Dr. Joel D. Green\footnote{Member of the "DUST, ICE AND GAS IN TIME (DIGIT) HERSCHEL KEY PROGRAM" TEAM}). In addition, other lines usually associated with PDR emission such as OI [63 $\mu$m and 145 $\mu$m] lines were also detected using both ISO and Herschel. Particularly, OI [63 $\mu$m] emission is centered in Elias Elias 29, as shown in \citet{Green2013} and \citet{RiviereMarichalar2016}, and is probably originated from the cavity walls illuminated by the central source given an almost face-on inclination \citep{Boogert2002, RochaPilling2015}. In this sense, these lines emission cannot be used to constrain the external irradiation on Elias 29 itself. 

\subsubsection{H$_2$CO emission as tracer of envelope temperature}
A study about externally heated protostellar cores in the Oph cloud was recently carried-out by \citet{Lindberg2017} using APEX 218 GHz observations of molecular emission (H$_2$CO and c-C$_3$H$_2$) as tracer of envelope temperature. They state that H$_2$CO emission is able to trace the temperature for a radius of R $>$ 2000 AU, whereas c-C$_3$H$_2$ originates at inner regions of envelope. In this way, the transition para-H$_2$CO [3$_{03}$-2$_{02}$] at 218.2222 GHz is reported, leading to an upper limit for rotational temperature of 109 K in Elias 29. The molecule c-C$_3$H$_2$, however, has not been observed. In this paper, the rotational temperature was recalculated using three transitions reported by \citet{Boogert2002} obtained with James Clerk Maxwell Telescope (JCMT) and National Radio Astronomy Observatory (NRAO) as shown in Table \ref{rotational}.

\begin{table*}
\begin{center}
\caption{Rotational parameters of para-H$_2$CO. Information extracted from \citet{Boogert2002} and \citet{Muller2001}\label{rotational}}
\begin{tabular}{ccccccccc}
\hline\hline
Molecule   & Transition  & $\nu$(GHz)  & E$_u$/k & A$_{ul}$(s$^{-1}$)  & g$_u$ & $\eta_{mb}$ & $\int T_{MB}dv$[km s$^{-1}$] & Telescope\\
\hline
p-H$_2$CO & 2$_{02}$-1$_{01}$ & 145.6030 & 10.48 & 7.8 $\times$ 10$^{-5}$ & 5 & 0.81 & 0.39 & NRAO\\
p-H$_2$CO & 3$_{03}$-2$_{02}$ & 218.2222 & 20.96 & 2.8 $\times$ 10$^{-4}$ & 7 & 0.68 & 0.82 & JCMT\\
p-H$_2$CO & 3$_{22}$-2$_{21}$ & 218.4756 & 68.09 & 1.6 $\times$ 10$^{-4}$ & 7 & 0.68 & $<$0.12 & JCMT\\
\hline
\end{tabular}
\end{center}
\end{table*}

Following \citet{Lindberg2017}, and assuming that molecules are in LTE, the upper-level column density ($N_u$) of formaldehyde was calculated using the following equations described in \citet{Goldsmith1999}:
\begin{equation}
N_u = \frac{8\pi k \nu^2}{h c^3 A_{ul} \eta_{mb}} \int T_{MB}dv
\end{equation}

\begin{equation}
\ln\left(\frac{N_u}{g_u} \right) = \ln N_{tot} - \ln Q(T_{rot}) - \frac{E_u}{kT_{rot}}
\end{equation}
where k, $\nu$, h, c are the obvious physical constants. $\int T_{MB}dv$ is the integrated intensity, $A_{ul}$ and $\eta_{mb}$ are the spontaneous Einstein coefficient and  beam efficiency, respectively. $g_u$ is the degeneracy of upper level, $N_{tot}$ is the total column density of formaldehyde, $Q(T_{rot})$ is the partition function at the rotational temperature $T_{rot}$ and $E_u$ is the energy at upper level. 

\begin{figure*}
\centering
\includegraphics[height=7cm]{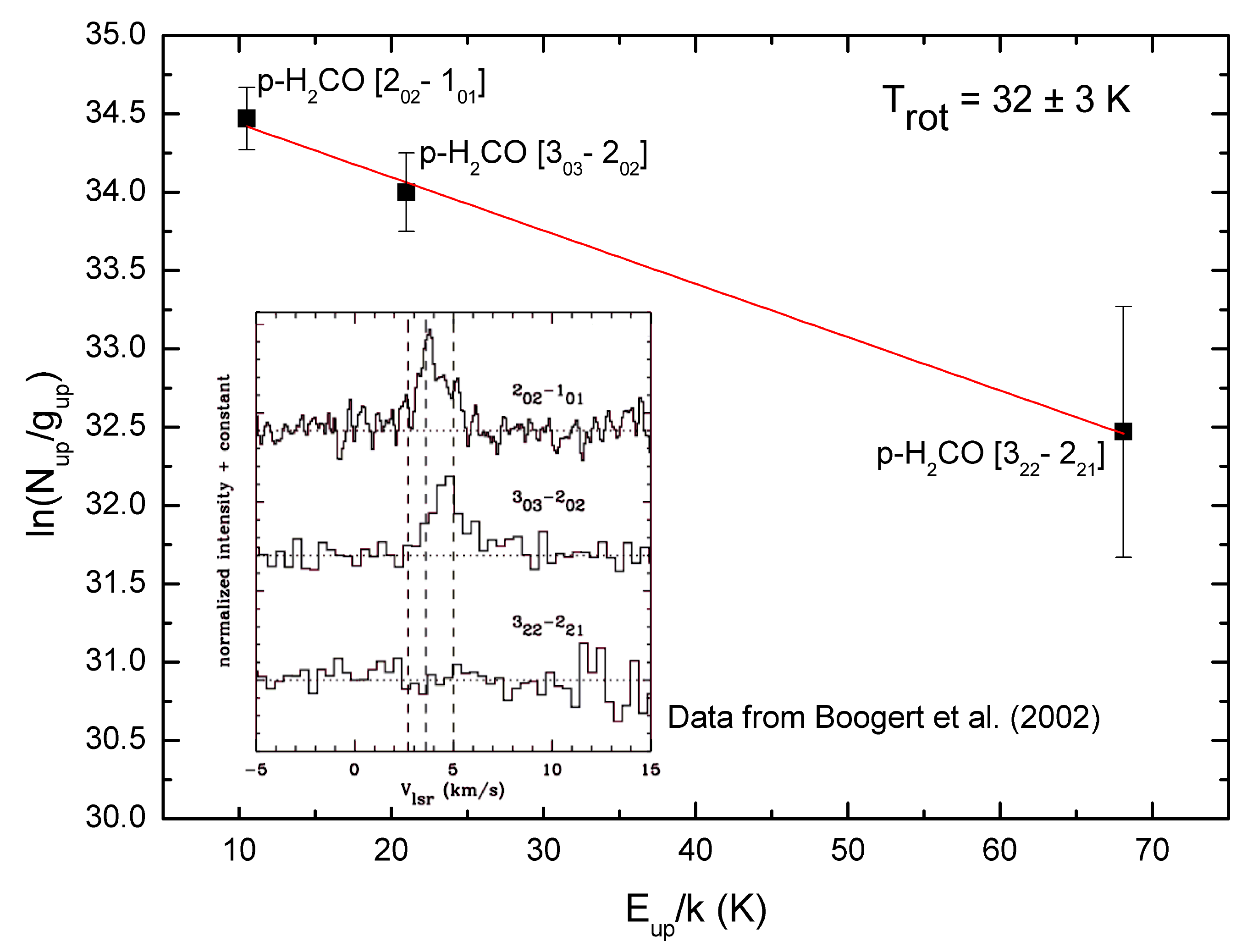}
\caption{Rotational diagram for three transitions of para-H$_2$CO molecule.}
\label{rotdiagram}
\end{figure*}

The temperature at outer envelope of Elias 29 calculated from the rotational diagram presented in Figure \ref{rotdiagram} is T$_{rot}$ = 32 $\pm$ 3 K. This seems to be a mean temperature at $\rho$ Oph cloud, once the range calculated in \citet{Lindberg2017} is between 20 - 50 K, excluding the values shown as upper limit.

\subsubsection{External irradiation constrained from flux extinction}
\begin{figure*}
\centering
\includegraphics[height=10cm]{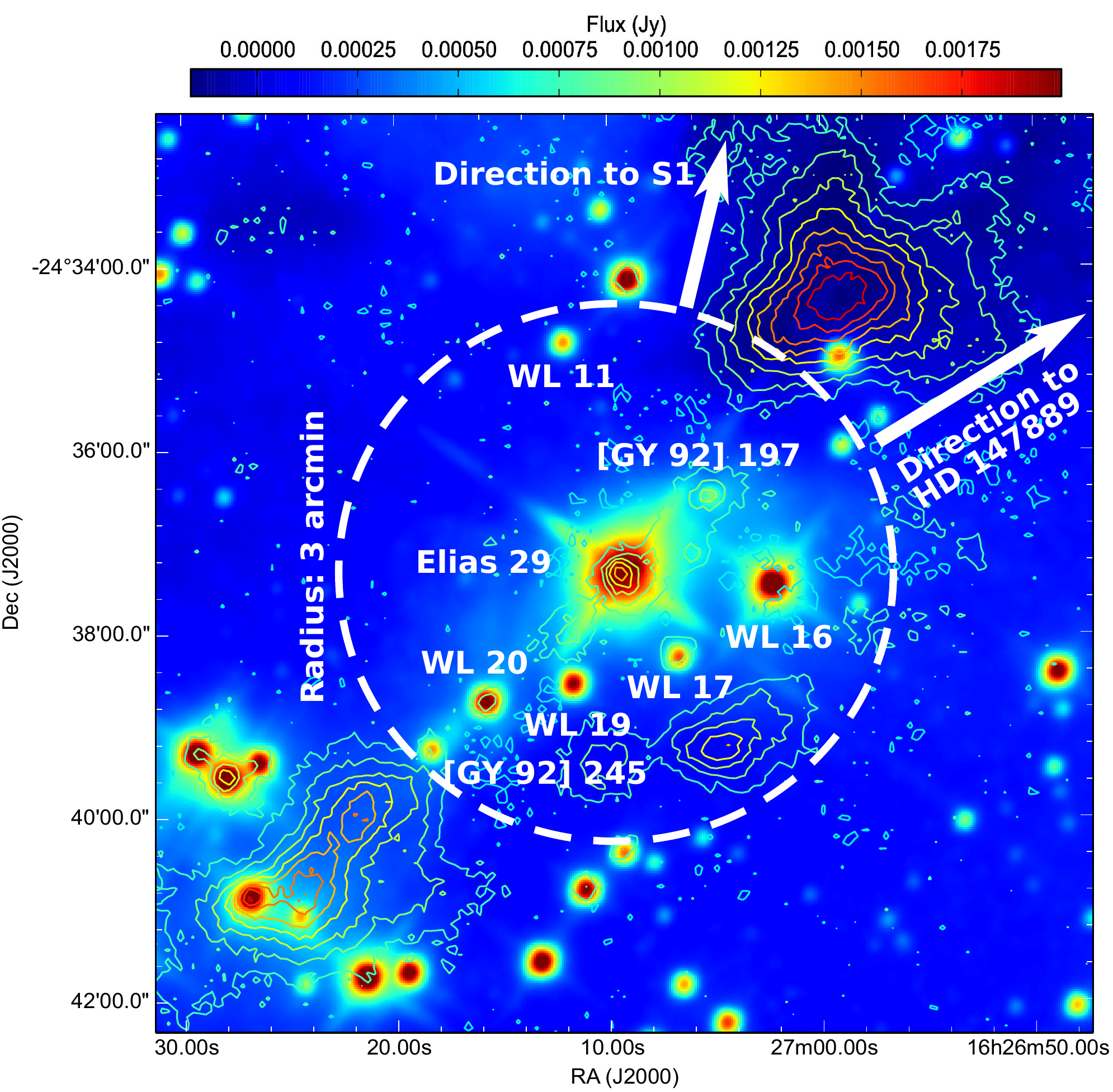}
\caption{WISE image centered in Elias 29 with FOV of 10 $\times$ 10 arcmin. The colors represent the flux in Jy. The contours are the dust emission in 850 $\mu$m collected from SUBARU/JCMT in Jy/pixel. The dashed circle limits a region with radius of 3 arcmin. The arrows indicate the direction to S1 and HD 147889.}
\label{wise}
\end{figure*}

Although B-type stars dominate the UV irradiation in $\rho$ Oph cloud, the closely Young Stellar Objects (YSOs) around Elias 29, are also potential sources of heating. Figure \ref{wise} shows a near-IR image (10 $\times$ 10 arcmin) obtained with WISE\footnote{irsa.ipac.caltech.edu/Mission/wise.html} in 3.4 $\mu$m centered in Elias 29. Contours indicate the cold dust emission in 850 $\mu$m obtained from SUBARU/James Clerk Maxwell Telescope (JCMT)\footnote{www.cadc-ccda.hia.nrc-cnrc.gc.ca/en/jcmt}. This image shows that, at least, 7 objects are distant of Elias 29 in a radius of 3 arcmin, and are listed in Table \ref{list}, as well as S1 and HD 147889.

\begin{table*}
\begin{center}
\caption{List and characteristics of the objects presented in Figure \ref{wise}.\label{list}}
\begin{tabular}{cccccccc}
\hline\hline
Object   & R.A. (J2000)  & DEC (J2000)  & Distance$^a$ & Relative distance$^j$  & L$_{\mathrm{bol}}$ & L$_{\mathrm{UV}}$ $^l$ &  $\overline{A_V^i}$ $^m$\\
         & (h m s)         & (d m s)    & (pc) & (pc) & (L$_{\odot}$) & (L$_{\odot}$) & (mag)\\
\hline
WL 11 & 16 27 12.131 & -24 34 49.14 & 128$^b$ & 6.5 $\times$ 10$^{-3}$ & 0.04$^e$ & 0.01 & 26.4\\
WL 16 & 16 27 02.340 & -24 37 27.20 & 125$^c$ & 1.1 $\times$ 10$^{-2}$ & 250.00$^c$ & 10.30 & 26.4\\
WL 17 & 16 27 06.776 & -24 38 15.20 & 137$^d$ & 1.4 $\times$ 10$^{-2}$ & 0.60$^e$ & 0.07 & 26.4\\
WL 19 & 16 27 11.776 & -24 38 32.02 & 123$^e$ & 1.2 $\times$ 10$^{-2}$ & 58.00$^e$ & 3.05 & 28.4\\
WL 20 & 16 27 15.730 & -24 38 43.70 & 110$^f$ & 8.3 $\times$ 10$^{-3}$ & 1.80$^e$ & 0.17 & 27.8\\
$\mathrm{[GY 92]}$ 197& 16 27 05.246 & -24 36 29.79 & 100$^g$ & 1.4 $\times$ 10$^{-2}$ & 0.15$^e$ & 0.02 & 25.9\\
$\mathrm{[GY 92]}$ 245& 16 27 18.380 & -24 39 14.68 & 105$^h$ & 6.0 $\times$ 10$^{-3}$ & 0.12$^e$ & 0.02 & 25.6\\
HD147889 & 16 25 24.317 & -24 27 56.57 & 120$^i$ & 7.0 $\times$ 10$^{-2}$ & 4500.00$^k$ & 116.00 & 18.2\\
S1 & 16 26 34.167 & -24 23 28.26 & 120$^i$ & 5.1 $\times$ 10$^{-2}$ & 1100.00$^k$ & 35.70 & 20.0\\
\hline
\multicolumn{8}{l}{$^a$ Distance from Earth; $^b$ \citep{Khanzadyan2004}; $^c$ \citep{Zhang2017}; $^d$ \citep{OrtizLeon2017};}\\
\multicolumn{8}{l}{$^e$ \citep{Bontemps2001}; $^f$ \citep{ResslerBarsony2001}; $^g$ \citep{Jorgensen2008};}\\
\multicolumn{8}{l}{$^h$ \citep{deGeus1989}; $^i$ \citep{Lindberg2017};}\\
\multicolumn{8}{l}{$^j$ Distance between each object in the Ophiuchi cloud:http://docs.astropy.org/en/stable/api/astropy.coordinates.SkyCoord.html}\\
\multicolumn{8}{l}{$^k$ \citep{Liseau1999}; $^l$ $\mathrm{log_{10}L_{UV} = 0.836 \times log_{10}L_{bol} - 1.78}$ - \citet{Lee2015}; $^m$ see text for details}\\
\end{tabular}
\end{center}
\end{table*}

\begin{figure*}
\centering
\includegraphics[height=6cm]{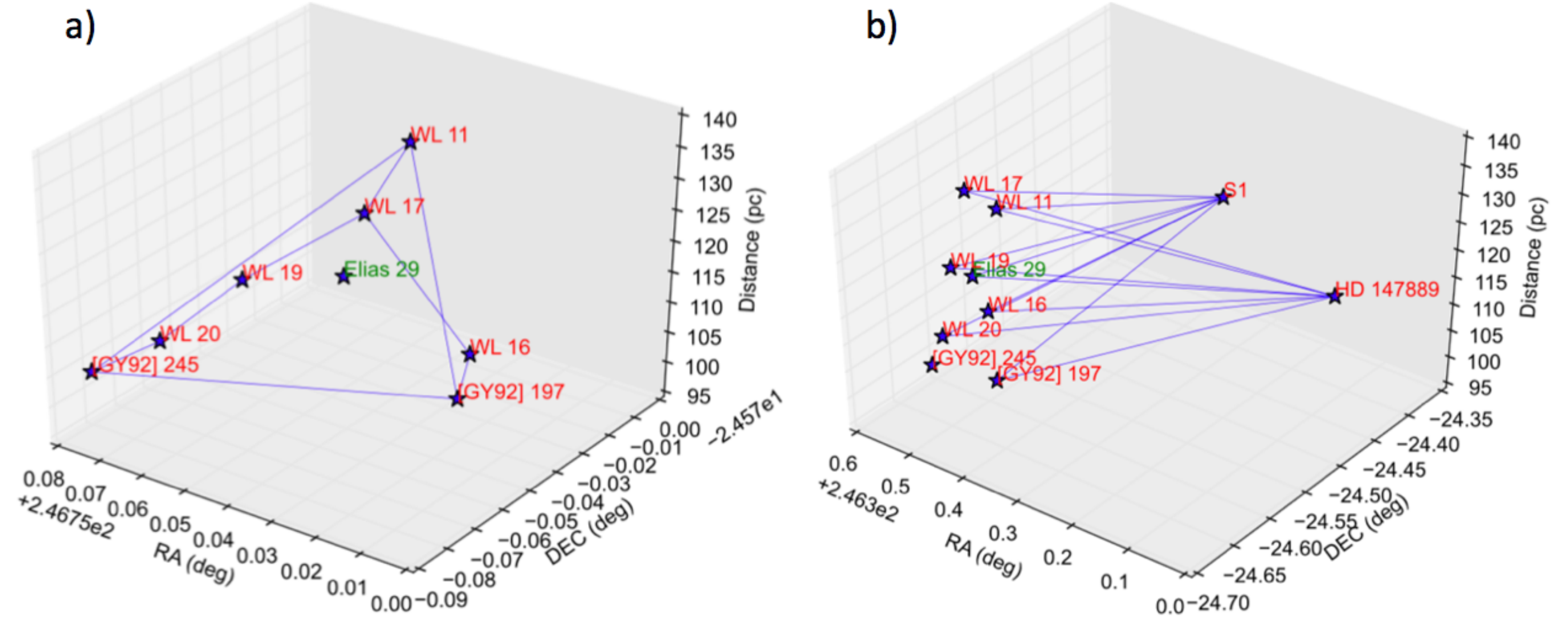}
\caption{Tree-dimensional spatial distribution of selected protostars in $\rho$ Oph cloud. (a) Stars symbols are objects inside the dashed circle in Figure \ref{wise}, and Elias 29 is highlight in green label. The connections among the objects shows that Elias 29 is embedded by UV radiation field. (b) Same of Panel (a), but now showing S1 and HD147889 B-type stars that dominate the radiation field in $\rho$ Oph cloud. The connections are only to guide the eyes.}
\label{3d}
\end{figure*}

Using the information provided in Table \ref{list}, Figure \ref{3d}a shows a 3D projection of each object inside the circle (Figure \ref{wise}) relative to Elias 29. The position of S1 and HD 147889 relative to Elias 29 is shown in Figure \ref{3d}b. Panels in Figure \ref{3d} are illustrative to show that is convenient to assume an isotropic external irradiation in simulations as will be shown in Section 2.2.

\begin{figure*}
\centering
\includegraphics[height=9.5cm]{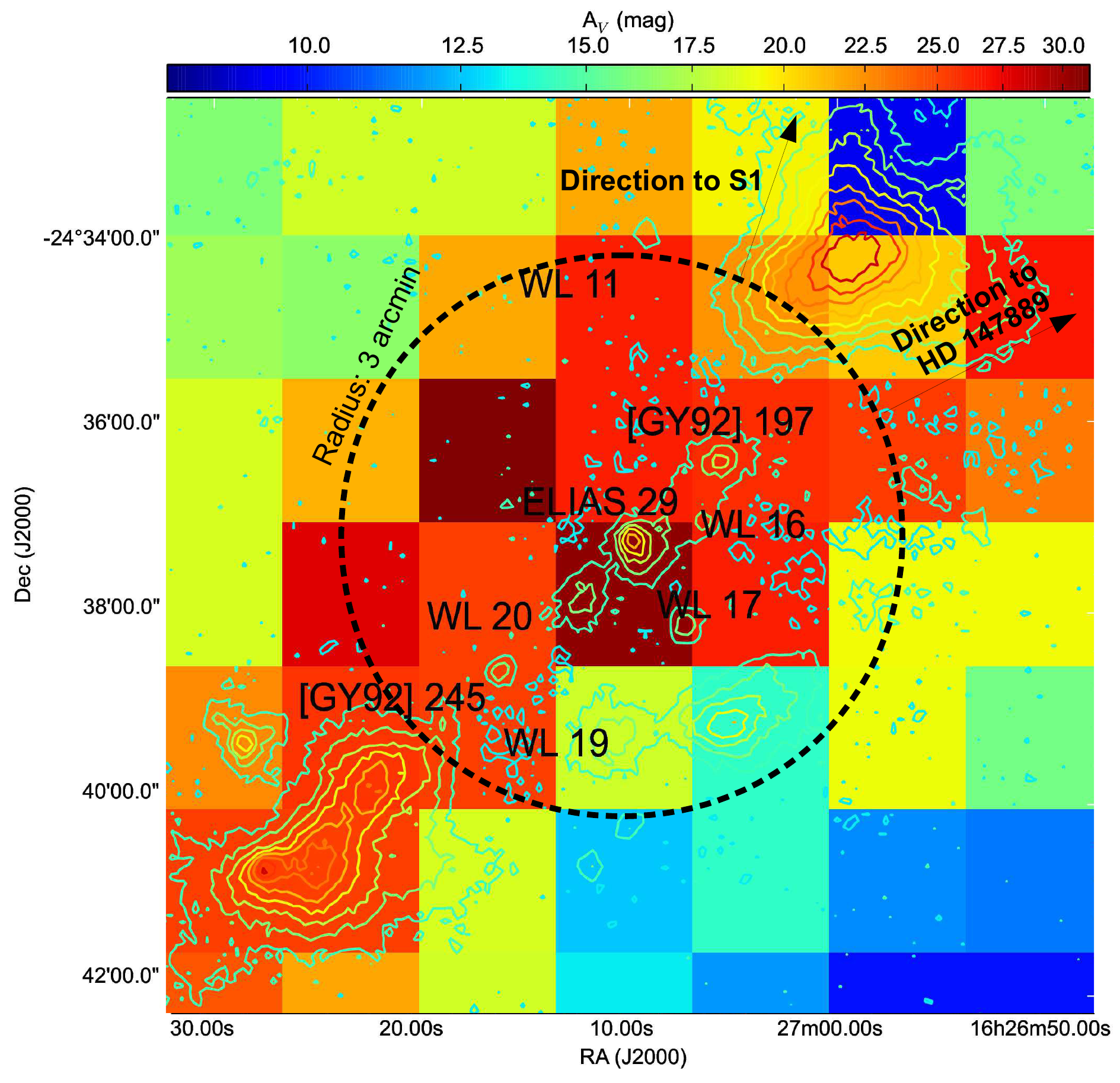}
\caption{Visual extinction (A$_V$) map centered in Elias 29 with FOV of 10 $\times$ 10 arcmin, obtained from COMPLETE (Coordinated Molecular Probe Line Extinction Thermal Emission) survey \citep{Ridge2006}. Contours, dashed circle and description are same of Figure \ref{wise}.}
\label{ext}
\end{figure*}

Once the environment around Elias 29 was characterized, the total incident flux was calculated, using the following equation:
\begin{equation}
F_{tot} = \sum_{i}^{n} \left(\frac{L_i^{\mathrm{UV}}}{4\pi R_i^2} \right) \exp\left[{\bar{\tau}_i^{\mathrm{UV}}} \right]
\end{equation}
where $F_{tot}$ is the total flux coming from neighbor objects (i to n) around Elias 29 as shown in Table \ref{list}, $L_i^{\mathrm{UV}}$ is the UV luminosity, $R_i$ the relative distance at $\rho$ Oph cloud. Figure \ref{ext} was used to constrain the UV extinction between each object and Elias 29, assuming the relation $\mathrm{A_{UV} \approx 2{A_V}}$ \citep{Fitzpatrick1999, Draine2003}. This figure shows a region of 10 $\times$ 10 arcmin of $\rho$ Oph containing the visual extinction ($\mathrm{A_V}$) map \citep{Ridge2006}. Were employed the following equations to obtain the mean ultraviolet optical depth ($\mathrm{\bar{\tau}_i^{\mathrm{UV}}}$) along the distance $R_i$:
\begin{equation}
\mathrm{\bar{\tau}_i^{\mathrm{UV}}} = \left(\frac{2}{m}\sum_{j}^{m} A_V^{ij} - A_V^{for}\right)\rho_d^{rel} \frac{R_i}{d_i}
\end{equation}
where $\frac{2}{m}\sum_{j}^{m} A_V^{ij}$ is the mean UV extinction calculated from Figure \ref{ext} along the pixels {\it j} to {\it m} between each object {\it i} and Elias 29, the visual extinction due to foreground clouds estimated to be 11 mag \citep{Boogert2002}, $\rho_d^{rel}$ is the local dust density relative to ISM, assumed to be 10$^5$ from Boogert et al., and $d_i$ is the distance of each object along the line of sight in parsec. 

Using this methodology, it was found that external irradiation interacting with Elias 29 is 44$\chi_{ISRF}$. In percentage terms, HD147889 roughly contributes with 93\% with the external UV field, whereas S1, WL 16 and others contribute with 5\%, 1.5\% and 0.5\%, respectively. 

\subsection{Continuum radiative transfer simulation including internal and external irradiation}
\begin{figure*}
\centering
\includegraphics[height=9.5cm]{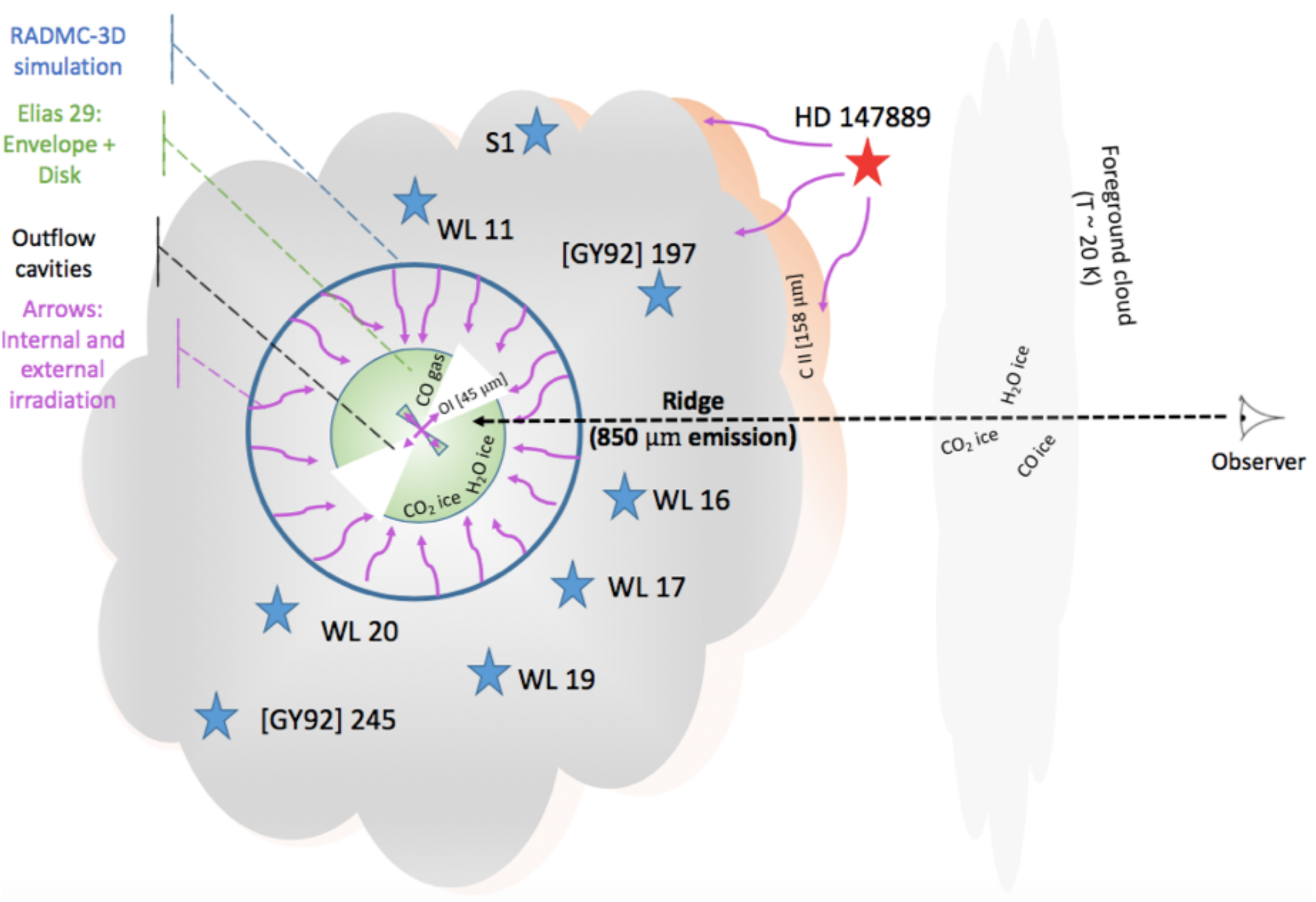}
\caption{Schematic illustration of Elias 29 environment. The disk inclination is 60$^{\circ}$ relative to observer. The internal protostar and external objects are source of photons in Elias 29, although HD 147889 dominates the external irradiation and might ionize the C atom. Figure not in scale.}
\label{ilus}
\end{figure*}

In order to understand how an external irradiation affects the survival of ices in Elias 29, the RADMC-3D code was used. It performs a three-dimensional Monte Carlo radiative transfer calculation considering an axisymmetric density profile. Specifically in this paper, the code was used to calculate the dust temperature and total flux in Elias 29, but no gas distribution was assumed. Figure \ref{ilus} shows an illustration of environment of Elias 29 and the region limited to RADMC-3D simulation. 

\subsubsection{Disk and envelope parameters}
The radiative transfer for internal irradiation was already simulated in Paper I, and the physical parameters are shown in Table \ref{params}. The disk and envelope density profile was ruled by using the following equations as described in Paper I:

\begin{table*}
\begin{center}
\caption{Physical parameters used in the radiative transfer simulation\label{params}}
\begin{tabular}{cccc}
\hline\hline
\multicolumn{4}{l}{\hspace{3cm} \citet{RochaPilling2015}}\\
Parameter   & Description  & Values  & Literature reference\\
\hline
R (R$_{\odot}$) & stellar radius & 5.7 & \citet{Miotello2014}\\
T (K) & Blackbody temperature & 4880 & \citet{Miotello2014}\\
L (L$_{\odot}$) & Luminosity & 16.5 & \citet{Miotello2014}\\
M$_d$ (M$_{\odot}$) & Disk mass & 0.003 & \citet{Lommen2008}\\
R$_d^{in}$ (AU) & Disk inner radius & 0.36$^a$ & ...\\
R$_d^{out}$ (AU) & Disk outer radius & 200 & \citet{Lommen2008}\\
M$_env$ (M$_{\odot}$) & Envelope mass & 0.028 & \citet{Lommen2008}\\
R$_{env}^{in}$ (AU) & Envelope inner radius & 0.36$^a$ & ...\\
R$_{env}^{out}$ (AU) & Envelope outer radius & 6000 & \citet{Motte1998}\\
$\theta$($^{\circ}$) & Cavity angle & 30 & \citet{Beckford2008}\\
d(pc) & Distance & 120 & \citet{Boogert2002}\\
\hline
\multicolumn{4}{l}{\hspace{4.0cm} This paper}\\
$\chi$ & Standard UV Field & 44 & see text\\
\hline
\multicolumn{4}{l}{$^a$ Value calculated using $R_{in} = R_{\star}(T_{\star}/T_{in})^2$ from \citet{Dullemond2010}}\\
\end{tabular}
\end{center}
\end{table*}

\begin{equation}
\rho_{disk}(r,\theta) = \frac{\Sigma_0\left(r/R_0 \right)^{-1}}{\sqrt{2\pi}H(r)} exp\left[-\frac{1}{2} \left(\frac{r\cos\theta}{H(r)} \right)^2 \right] 
\end{equation}
\begin{equation}
\rho_{env} = \rho_0\left(\frac{R_{out}}{r} \right)
\end{equation}
where $\theta$ is the angle from the axis of symmetry; $\Sigma_0$ is the surface density at outer radius $R_0$ and H(r) is the disk scale height, given by $H(r)=r(H_0/R_0)(r/R_0)^{2/7}$, that is defined from a self-irradiated passive disk proposed by \citet{Chiang1997}. Further, $\rho_0$ is the density at outer radius of the envelope R$_{out}$. The equation (5) was also used in models by \citet{Pontoppidan2005,Lommen2008,Crapsi2008}, whereas the equation (6) was modified from \citet{Lommen2008} to characterize a static envelope as used in \citet{Pontoppidan2005}.

It is worth to note that disk dimension in Elias 29 present different values in literature. \citet{Boogert2002} has fixed the disk size in 500 AU, whereas \citet{Miotello2014} propose a size between 15 - 200 AU by considering an optically thick or thin disk, respectively. As claimed by authors, the first case is unrealistic because the dust properties cannot be addressed. In the second one, the disk should be populated by cm-sized pebbles. \citet{Huelamo2005} supports the idea that Elias 29 disk is around 200 AU due to direct image in K band obtained with VLT-ISAAC. This size is also employed in \citet{Lommen2008} as a fiducial limit.

\subsubsection{Dust model}
\begin{figure*}
\centering
\includegraphics[height=11cm]{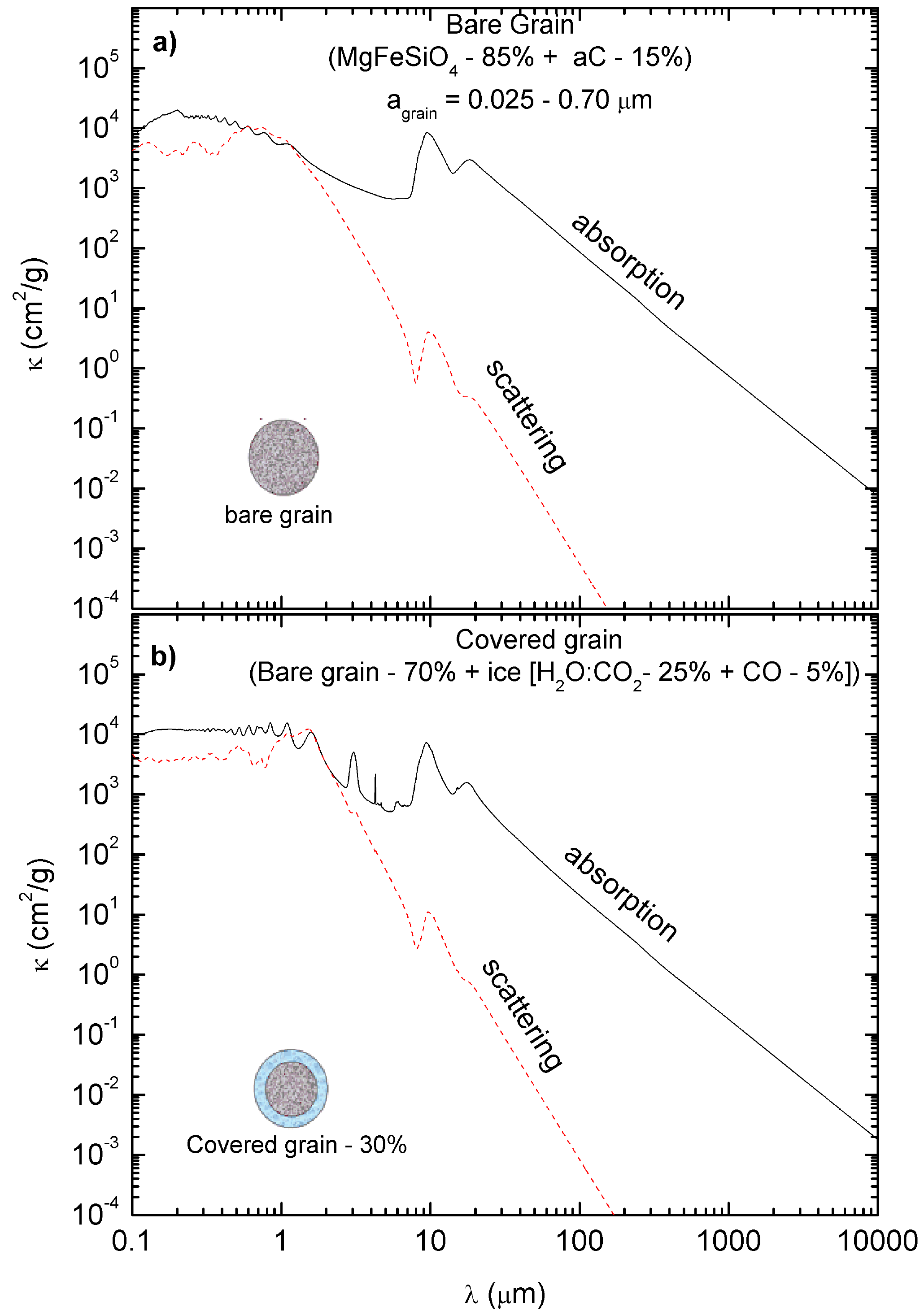}
\caption{Absorption and scattering opacities of interstellar grains, assuming a MNR size distribution. Panel (a) shows the opacities for bare grains, whereas Panel (b) shows the opacities of ice-covered grains.}
\label{opac}
\end{figure*}

The dust model used in simulations combines bare and covered grains ruled by a MRN size distribution \citep{Mathis1977} ranging between 0.025 - 0.70 $\micron$, constrained from \citet{Weingartner2001} and \citet{Beckford2008}. Bare grains have been used in warm and hot regions (150 - 1500 K), whereas covered grains in cold regions ($<$150 K) as seen in \citet{Lommen2008}. The bare grains are composed by magnesium iron silicate (MgFeSiO$_4$)\footnote{http://www.astro.uni-jena.de/Laboratory/OCDB/data/silicate/amorph/olmg50.lnk} mixed with amorphous carbon\footnote{ http://www.astro.uni-jena.de/Laboratory/OCDB/data/carbon/cel800.lnk}, at proportion 85\% and 15\%, respectively. The covered grains, on the other hand, are formed by a dust core (bare grain) and an ice mantle at proportion 70\% for dust and 30\% for ice, which is made of CO (5\%) and of H$_2$O:CO$_2$ (25\%) processed by radiation \citep{Rocha2017} in order to fit the chemical evolution in near to mid-IR as shown in Paper I. The dust opacities are shown in Figure \ref{opac}a and \ref{opac}b.

\subsubsection{External irradiation}
The RADMC-3D code simulated an isotropic ISRF by considering an external sphere with the same size of spatial grid as source of photons. In order to perform the Monte Carlo simulation, photon packages are launched from the sphere inward with the following flux:
\begin{equation}
F_{ISRF}^{UV} = \pi W_{dil} \int_{\Omega}\int_{\mathrm{91 nm}}^{\mathrm{205 nm}} B_{\lambda}(20000K) d\lambda d\Omega
\end{equation}
where $B_{\lambda}$ is a blackbody highly diluted by the factor of $W_{dil} = 8.4 \times 10^{-11}$ by assuming an external radiation field of 44$\chi_{ISRF}$ as it is described in Section 2.1.2. 

\section{Results}
\subsection{UV – centimeter SED and near-IR image}
The 0.1 - 100000 $\mu$m Spectral Energy Distribution (SED) of Elias 29 was modeled using RADMC-3D by considering four models, taking into account different values of external flux. The Models 1 - 3 adopt 0$\chi_{ISRF}$ (T$_d$ $<$ 20 K), 100$\chi_{ISRF}$ (T$_d$ = 35 K) and 44$\chi_{ISRF}$ (T$_d$ = 30 K), respectively. The best model, called Model 4, is the Model 3 in addition to graybody emission from foreground cloud. Such model reproduces both near-IR and centimeter emission taken from \citet{Lommen2008} and \citet{Miotello2014}. 

\begin{figure*}
\centering
\includegraphics[height=16cm]{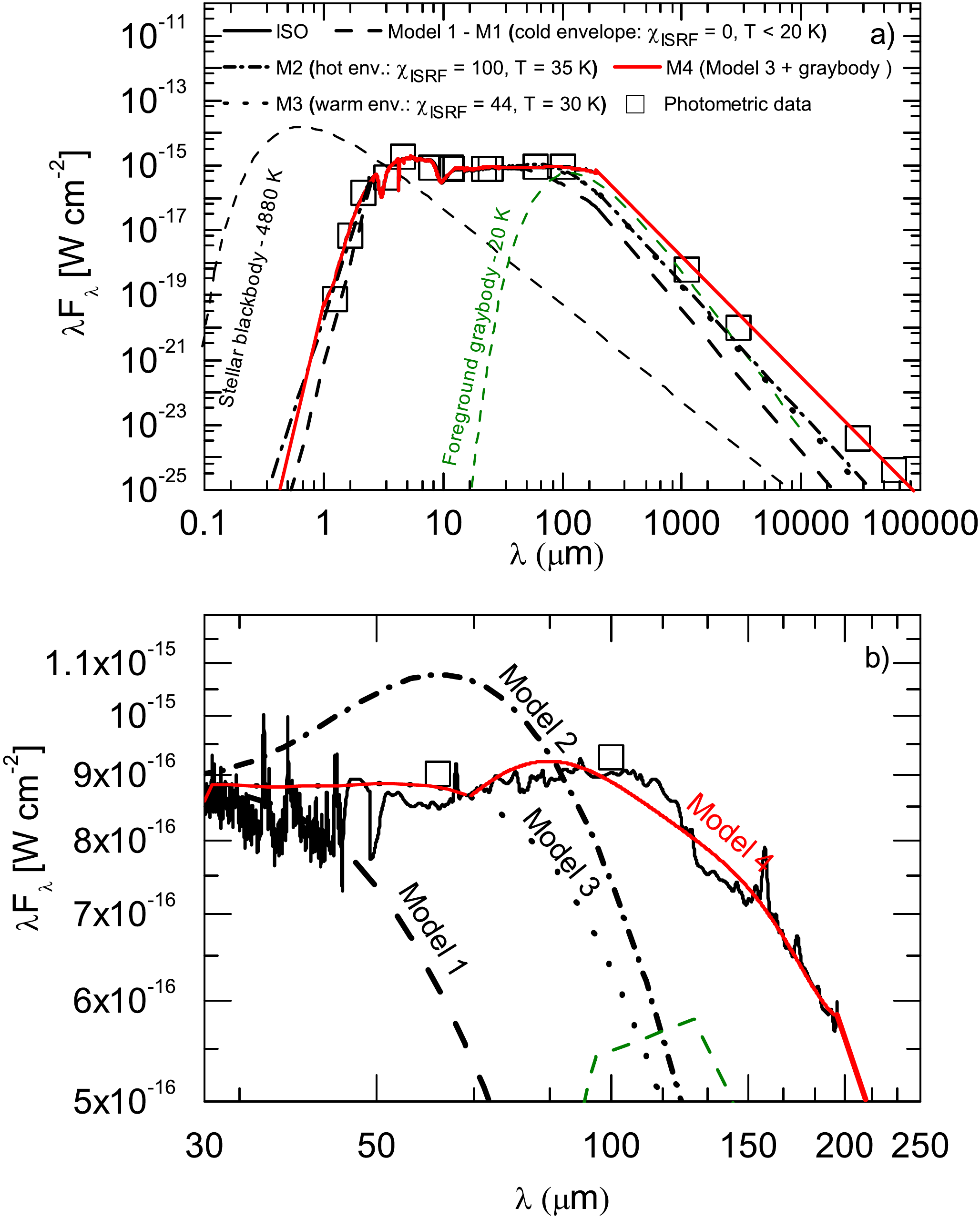}
\caption{SED of Elias 29 assuming an internal and external irradiation. (a) 0.1 - 100000 $\mu$m SED highlighting the Models 1 - 4 and the photometric data taken from literature (2MASS, WISE, Spitzer, IRAS, SMA and ATCA). (b) Zoom of panel (a) showing details of the Models 1 - 4.}
\label{sed}
\end{figure*}

Figure \ref{sed}a shows the whole SED of Elias 29, emphasizing the models and observational data from ISO and photometric data. Figure \ref{sed}b shows the SED between 30 - 250 $\mu$m to highlight the contribution of different values of external irradiation between mid and far-IR. One can observe by Model 1 that disk+envelope alone cannot reproduce emission for $\lambda$ $>$ 30 $\mu$m, which means that another component is contributing for increasing the emission in mid-IR. \citet{Boogert2002} claims that ridge temperature is around 15$\pm$5 K, which is low to justify the observed emission in the spectrum. They also reports other two components toward Elias 29 with velocities 2.7 km s$^{-1}$ and 3.8 km s$^{-1}$ with temperatures around 15$\pm$5 K constrained from intensity ratio of C$^{18}$O 1-0/3-2. Nevertheless, this value still low to explain such emission. To overcome this problem, they assume that second component with 3.8 km s$^{-1}$ should have a temperature between 20 - 40 K. Previous sections in this paper, however, have shown that Elias 29 is surrounded by a strong UV emission from B-type objects, and the temperature in its envelope should be around 30 K constrained from H$_2$CO rotational lines. In this way, there is no reason to consider another foreground component with high temperature, but rather, that Elias 29 contains a warm envelope.

The Models 2 and 3 support the idea of an external irradiation not larger than 50$\chi_{ISRF}$ in Elias 29. Once such increasing flux comes from background emission, it is important to calculate the spectrum using the same spectral aperture of ISO LWS ($\lambda$ $>$ 45 $\mu$m) that is 86 arcsec \citep{Clegg1996}. Nevertheless, the long-wavelength ($\lambda$ $>$ 100 $\mu$m) spectrum is not reproduced is such scenario, and the contribution of a warm foreground cloud should be considered in this case. In this way, Model 4 is the best model to reproduce the Elias 29 spectrum from UV to centimeter regime.

The structure of Elias 29 was probed by \citet{Brandner2000} using the ISAAC (Infrared Spectrometer and Array Camera) coupled to Very Large Telescope (VLT) in H (1.63 $\mu$m) and K (2.16 $\mu$m) band in polarimetric mode. The authors report a bipolar-nebula structure in K band in FOV of 1 arcmin. \citet{Huelamo2005}, in addition, probed the inner structure of Elias 29 in FOV of 3 arcsec using NACO/VLT polarimetric differential imaging. A dark lane is detected in the NE-SW direction, indicating the presence of protostellar disk with radius larger than 180 AU.

\begin{figure*}
\centering
\includegraphics[height=5cm]{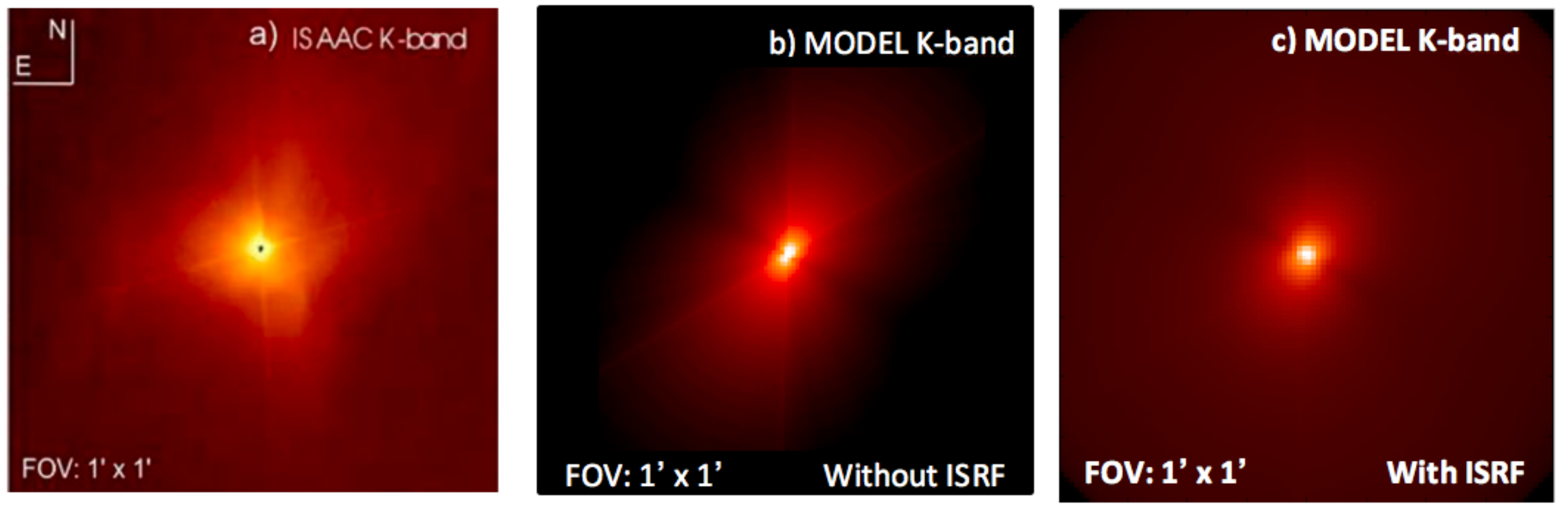}
\caption{Comparison among observation and model of Elias 29 as seen in K-band (2.16 m). Panel (a) shows Elias 29 as observed with ISAAC/VLT in polarimetric mode (adapted from \citet{Huelamo2005}). Panels (b) and (c) show Elias 29 simulated with RADMC-3D assuming a scenario without and with external irradiation, respectively. No polarization is calculated in the models. }
\label{imageK}
\end{figure*}

In Paper I the modelling of Elias 29 in K band considering only the internal source of irradiation is presented. In this paper, however, the external irradiation is included in the simulation, according to Model 4. This methodology was employed in \citet{Gramajo2010} by combine SED + image approached by eye to obtain more reliable results. The real image of Elias 29 is shown in Figure \ref{imageK}a \citep{Huelamo2005}, whereas the synthetic image in Figure \ref{imageK}c. Panel b shows the modelling without external irradiation as considered in Model 1. The brightness profile that highlights the bipolar nebulosity in real image (Panel a), compared with Models 1 and 4 comes from scattered polarized light, once such image was observed in polarimetric mode.

\subsection{Temperature distribution}
\begin{figure*}
\centering
\includegraphics[height=10cm]{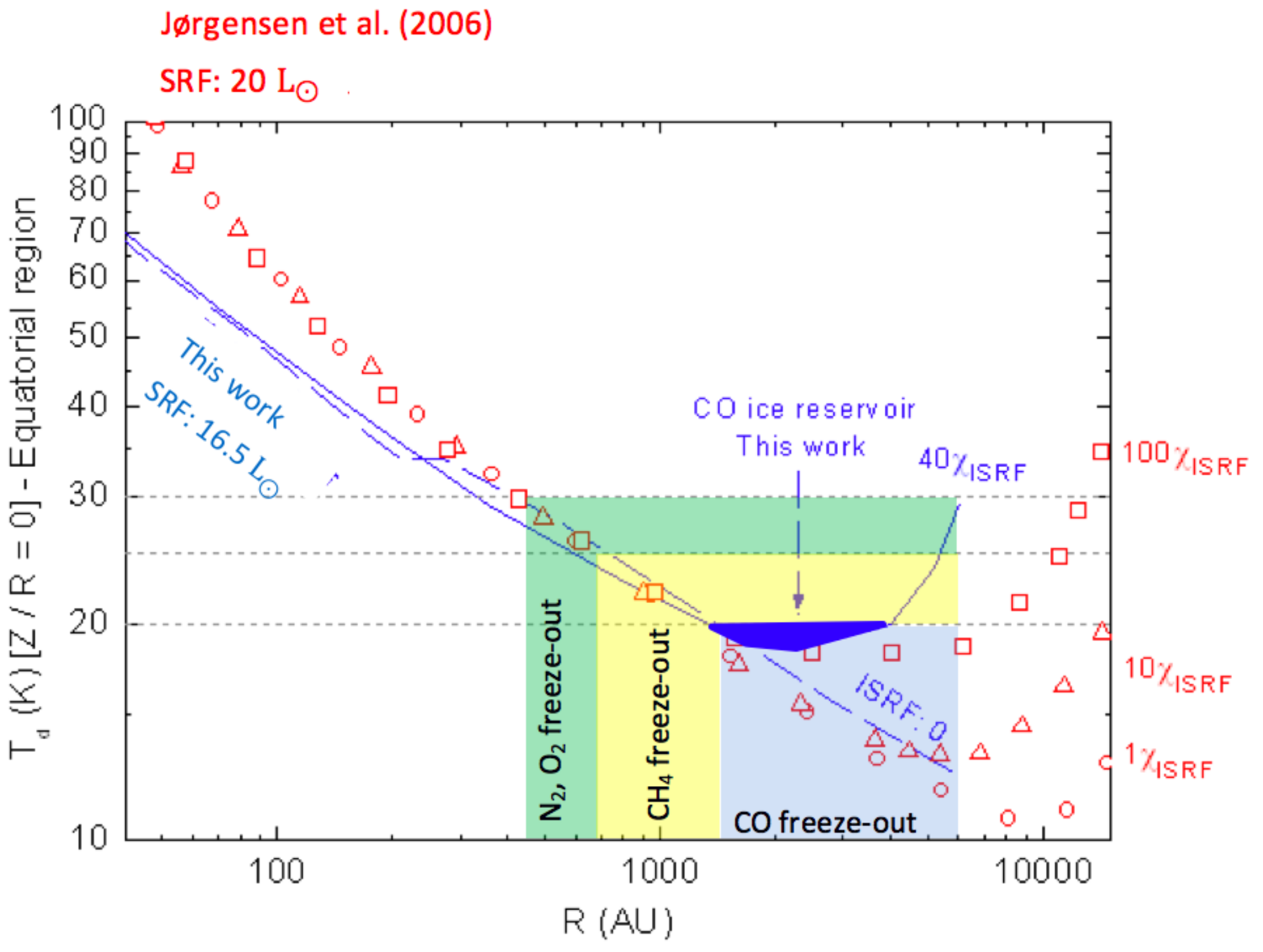}
\caption{1D dust temperature profiles as function of radius. Dashed and solid blue lines are relative to Elias 29 Models 1 and 4, respectively. Red symbols concern to models described in \citet{Jorgensen2006} for $\chi/\chi_{ISRF}$ of 1, 10 and 100. The freeze-out regions of volatile species are also indicated by the colored regions Dark blue region highlight the CO ice reservoir in Elias 29. }
\label{temp}
\end{figure*}

The routine {\it mctherm} in RADMC-3D was used to calculate the dust temperature of Elias 29. Figure \ref{temp} compares the 1D temperature profiles for Model 1 and Model 4 with the work from \citet{Jorgensen2006}, that illustrates the competition between the internal and external heating for a  envelope. The numerical density at outer radius is  for a range between 50 - 15 000 AU.

The models in J{\o}rgensen et al. assumes an internal source with 20$\mathrm{L_{\odot}}$ and different $\chi_{ISRF}$ of 1, 10 and 100. One can observe that even a typical interstellar radiation (1$\chi_{ISRF}$) might heat the outer envelope by a few degrees. However, only after 10$\chi_{ISRF}$ the temperature is increased above 20 K and might present important consequences for chemistry once the CO-ice desorbs around 18 - 20 K \citep{Collings2004}. From Elias 29 models it is possible observe that an external irradiation of 44$\chi_{ISRF}$ increases the outer envelope up to 30 K and between 1500 - 4000 AU, forms a CO ice reservoir. Such temperature profile, including an external irradiation, was also described in \citet{Launhardt2013} for starless and protostellar cores. They show that variation of observed mean outer dust temperature ($\mathrm{\Delta T_d^{outer}}$) for a dataset of 12 globules is around 5 K, by considering $\chi$ of 5 and $\mathrm{A_V}$ = 4 mag. Such result is also similar to J{\o}rgensen et al. models, and support the idea that a strong external irradiation is necessary to drive significant chemical changes in the envelope.  

\begin{figure*}
\centering
\includegraphics[height=14cm]{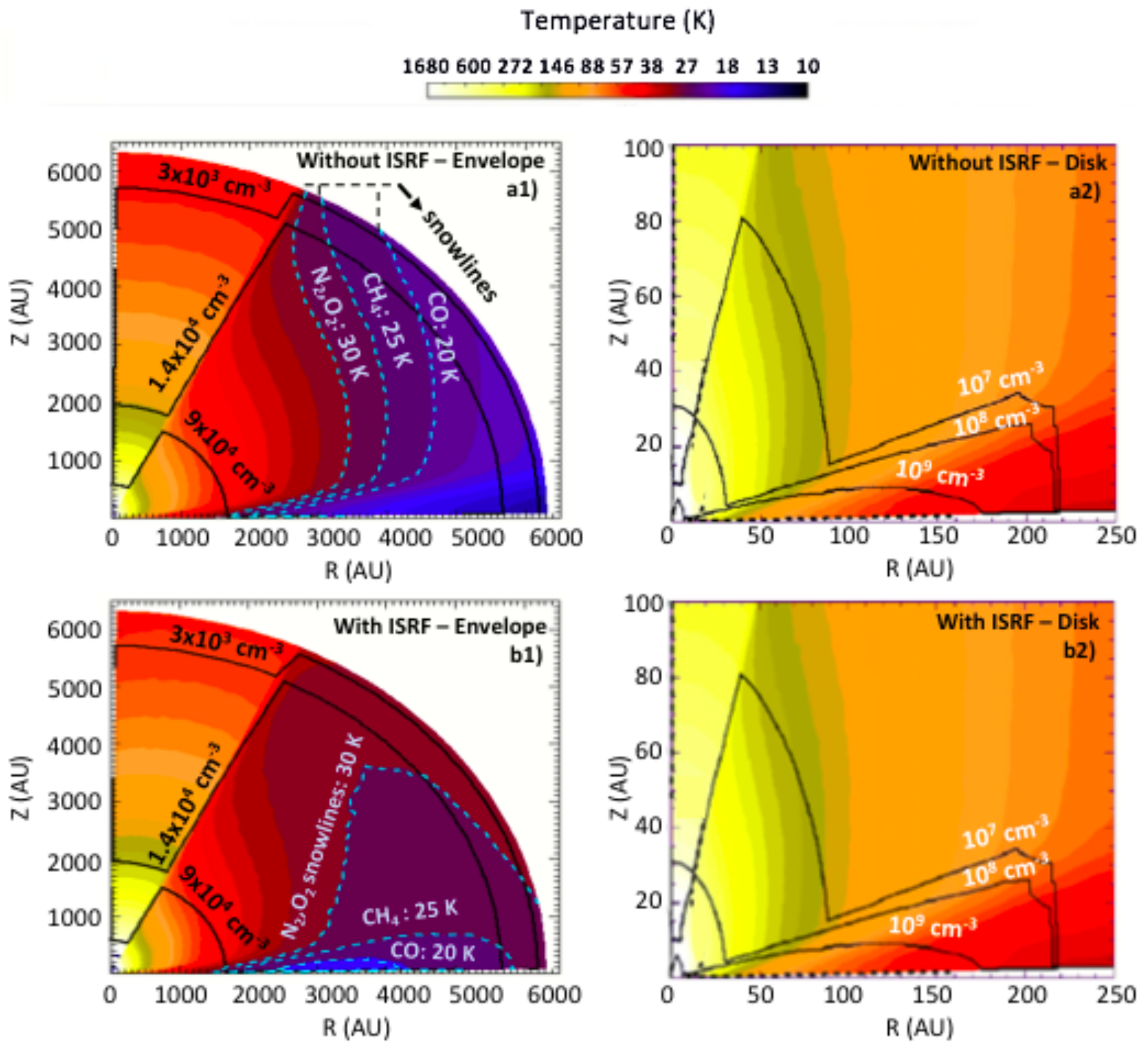}
\caption{Temperature map of Elias 29 including and excluding the ISRF (model 4 and 1, respectively). Solid lines represent the H$_2$ densities, and dashed lines the snowlines for CO (20 K), CH$_4$ (25 K) and N$_2$, O$_2$ (30 K). A cold toroidal-shape region in the envelope can be easily observed in panel b1. Nothing change in the disk region.}
\label{temp2D}
\end{figure*}

Figure \ref{temp2D} shows the 2D temperature profile of Elias 29, emphasizing the disk and envelope. The scenario without external irradiation, as shown in Panels a1 and a2, was already described in details in Paper I, in which only the internal protostar is a source of heating. In this paper, however, it is presented a new discussion considering an external source of UV photons as seen in Panels b1 and b2. By comparing both panels, one can observe that external irradiation might redefine the snowline position of volatile species such as N$_2$- and O$_2$-ice, CH$_4$-ice and CO-ice that desorbs at 30 K, 25 K and 20 K, respectively according to laboratory experiments detailed in \citet{Collings2004}. In the presence of an external UV field, the snowlines changes from layered structure to a toroidal-shaped distribution. In spite they are volatile species, they might survive in low abundance up to temperature of 150 K, if they are trapped in a H$_2$O-ice matrix that represents a most realist scenario.

CO-ice has an important role for the chemistry of complex in solid-phase, and many laboratory experiments have shown that a repeated hydrogenation of the adsorbed CO on dust grain lead to the formation of H$_2$CO and CH$_3$OH \citep{Watanabe2002, Watanabe2003, Fuchs2009}. Nevertheless, the efficiency to form methanol ice by this pathway is highly dependent of the temperature \citep{Watanabe2003}. The authors claim that maximum yield of CH$_3$OH ice formation is reached for temperatures between 10 and 15 K, whereas at 20 K the abundance of this ice is significantly smaller. Further, Monte Carlo simulations from \citet{Cuppen2009} have demonstrated that in regions with densities of n $\geq$ 10$^5$ cm$^{-3}$ and a large H/CO ratio at low temperature, both H$_2$CO and CH$_3$OH ices are formed on a short timescale, which suggests that methanol ice can be present since the first stages of star-formation process. Another mechanism to form CH$_3$OH ice was presented in \citet{Bergner2017}, in which the authors employed an insertion of O atoms into CH$_4$-ice for temperatures below 25 K.

 The gas abundance of Elias 29 was reported in \citet{Boogert2000}, showing that CO is the most abundant gas, followed by H$_2$O and CO$_2$, with gas/solid ratios around 53, 0.23 and 0.011 respectively. Rotational diagrams for CO-gas indicate two populations as detailed in \citet{Boogert2002}: (i) hot-CO at 1100 K and N$_{hot}$(CO) = 2 $\times$ 10$^{18}$ cm$^{-2}$, (ii) cold-CO at 90 K and N$_{cold}$(CO) = (16 $\pm$ 10) $\times$ 10$^{18}$ cm$^{-2}$, which indicates that CO-gas is present in Elias 29 itself. H$_2$O and OH are also placed in Elias 29, once the rotational diagrams trace temperatures of 379 and 230 K, respectively. 

H$_2$CO and CH$_3$OH was also observed in the gas-phase toward Elias 29 as indicated in \citet{Boogert2002}. The $V_{LSR}$ of both molecules indicates that p-H$_2$CO transitions above 145 GHz belongs to outer envelope of Elias 29, whereas the transition 2 - 1 triplet of CH$_3$OH at 96 GHz belongs to foreground clouds. Once any other large or complex molecule was observed in Elias 29 direction, this could be an indicative that they are confined by snowlines presented in Figure 11 and are not observed due to inclination of 60$^{\circ}$.

Figures \ref{temp2D}a2 and \ref{temp2D}b2 emphasize the disk region in both models. One can observe that dust temperature still the same, and external irradiation do not have influence for R $<$ 250 AU. Even FUV photons coming through the low density cavity are extincted in a radius R $>$ 4500 AU in this model due to small dust grains. The snowlines of volatile species such as N$_2$, O$_2$ and CH$_4$ are not shown, once the outer disk present a temperature around 70 K. Even in this situation, volatile species might be present if trapped in non-volatile ice, as for example H$_2$O or CO$_2$.

\section{Conclusions}
In this paper, the interplay between external UV and stellar radiation field with the survival of astrophysical ices was addressed. Specifically, it is introduced a discussion about the formation of a toroidal-shaped region of volatile compounds limited by their snowlines at the outer envelope of Elias 29. The conclusions are summarized below:

\begin{enumerate}
\item The outer envelope of Elias 29 is more warm (T $\sim$ 30 K) than what is expected for protostars illuminated only by a central source of photons. Such temperature is justified by the presence of many YSOs around Elias 29 distant only by 3 arcmin, as well as by two brightness B-type stars (S1 and HD 147889). Comparatively, HD 147889 contributes in 93\% the FUV emission at $\rho$ Oph cloud, and produces an external irradiation of around 44$\chi_{ISRF}$ in Elias 29. In such scenario, the background emission observed in K-band is roughly reproduced.

\item In the Elias 29 environment, the FUV external irradiation might penetrate in the envelope and redefine the snowline position of volatile molecules such as N$_2$, O$_2$, CH$_4$ and CO to a toroidal-shaped form. In that case, the maximum abundance of CO ice is confined to a small region of envelope, and consequently, the abundance of complex molecules as result of hydrogenation and oxygenation mechanisms is expected to be severely reduced. To address the impact of external irradiation on the survival of COMs, this model will be used as template into the Physico-Chemical ProDiMo code \citep{Woitke2009, Kamp2010, Thi2011, Woitke2016, Kamp2017} in a forthcoming paper. 

\item Although the scheme of a toroidal region of volatile compounds, such as found in Elias 29, is exciting, this might not be a global scenario, once the closest YSOs does not contribute much to increase the ISRF (roughly 2\%). Nevertheless, other regions such as CrA and  $\rho$ Oph A, are also dominated by external irradiation, and a comprehensive modeling of the chemistry in these objects will improve the knowledge about the chemical heritage in environments externally irradiated.  
\end{enumerate}

\section*{Acknowledgements}
The authors thank FAPESP (PD 2015/10492-3) for the financial support of this work and also FVE/UNIVAP and CAPES. We acknowledge the positive comments from anonymous referee that highly improved this manuscript, as well as the private communications with Dr. Joel D. Green about Herschel observations toward $\rho$ Oph cloud. Finally, we thank all the comments of Dr. Peter Woitke about this paper.




\bibliographystyle{mnras}
\bibliography{References} 


\bsp	
\label{lastpage}
\end{document}